\documentclass[10pt,twocolumn,english,conference]{IEEEtran}
\usepackage[T1]{fontenc}
\usepackage[latin1]{inputenc}
\usepackage{geometry}
\usepackage{subfigure}
\usepackage{amsmath}
\usepackage{graphicx}
\usepackage{amssymb}

\makeatletter
 \newtheorem{thm}{Theorem}
 \newtheorem{remrk}{Remark}
 \newtheorem{lemma}{Lemma}

\usepackage{cite}
\usepackage{bm}
\usepackage{bbm}

\usepackage{babel}
\makeatother
\begin{document}

\title{Performance Analysis of CDMA Signature Optimization with Finite Rate
Feedback}

\author{\authorblockN{Wei Dai$^{\dagger}$, Youjian Liu$^{\dagger}$, Brian Rider$^{\ddagger}$} \authorblockA{\\$^{\dagger}$Dept. of Electrical and Computer Eng., $^{\ddagger}$Math Department\\ University of Colorado at Boulder\\ Boulder CO 80303, USA\\ Email: dai@colorado.edu, eugeneliu@ieee.org, brider@euclid.colorado.edu} }

\maketitle
\begin{abstract}
We analyze the performance of CDMA signature optimization with finite
rate feedback. For a particular user, the receiver selects a signature
vector from a signature codebook to avoid the interference from other
users, and feeds the corresponding index back to this user through
a finite rate and error-free feedback link. We assume the codebook
is randomly constructed where the entries are independent and isotropically
distributed. It has been shown that the randomly constructed codebook
is asymptotically optimal. In this paper, we consider two types of
signature selection criteria. One is to select the signature vector
that minimizes the interference from other users. The other one is
to select the signature vector to match the weakest interference directions.
By letting the processing gain, number of users and feedback bits
approach infinity with fixed ratios, we derive the exact asymptotic
formulas to calculate the average interference for both criteria.
Our simulations demonstrate the theoretical formulas. The analysis
can be extended to evaluate the signal-to-interference plus noise
ratio performance for both match filter and linear minimum mean-square
error receivers. 
\end{abstract}

\section{\label{sec:Introduction}Introduction}


In a direct-sequence code-division multiple access (DS-CDMA) system,
the performance is mainly limited by the interference among users.
To minimize the interference, every particular user wants to select
a signature vector from a signature codebook to avoid the interference
from other users. In this paper, we assume that the receiver (base
station) has the perfect information of the signatures. It selects
a signature for a particular user according to some criterion, and
feeds the corresponding index to this user through a feedback link.
We also assume that the feedback link is error-free and rate limited.
Due to the finite feedback rate, there is a performance degradation
compared to the infinite feedback rate case. We are interested in
quantifying the effect of finite rate feedback. 

This problem has been studied in \cite{Honig_IT05_CDMA_signature_optimization_finite_feedback}.
A randomly constructed signature codebook is assumed in \cite{Honig_IT05_CDMA_signature_optimization_finite_feedback}
where the codebook entries are independent and isotropically distributed.
The interference signature matrix is assumed to have independent and
identically distributed (i.i.d.) Gaussian elements. A particular user
chooses the signature vector from the signature codebook to maximize
signal-to-interference plus noise ratio (SINR). For the matched filter
receiver, this criterion is equivalent to select the signature to
minimize the interference. In \cite{Honig_IT05_CDMA_signature_optimization_finite_feedback},
an asymptotic lower bound is given on the average interference.

The main contribution of this paper is to derive the exact performance
limit. In this paper, we use the average interference as a performance
measure, which is independent of specific receivers and applications.
We consider two signature selection criteria. One is to minimize the
interference from other users, same as the one in \cite{Honig_IT05_CDMA_signature_optimization_finite_feedback}.
The other one is more intuitive. We select the signature vector to
match the weakest interference directions, or equivalently, to be
as orthogonal as possible to the strong interference directions. To
analyze the corresponding performance, we let the processing gain,
number of users and feedback bits approach infinity simultaneously
with fixed ratios. By asymptotic analysis, we derive lower bounds
and upper bounds on the average interference for both criteria. For
each criteria, the asymptotic upper bound meets the asymptotic lower
bound. Therefore, these bounds provide the exact performance limit.
The corresponding analysis can be extended to evaluate the SINR performance
for both match filter and linear minimum mean-square error (MMSE)
receivers.

\section{\label{sec:System-Model}System Model}


In a sampled discrete-time symbol-synchronous DS-CDMA system, the
received vector can be written as 

\[
\mathbf{Y}=\sum_{j=1}^{m}B_{j}\mathbf{s}_{j}+\mathbf{W},\]
where $B_{j}\in\mathbb{C}$ and $\mathbf{s}_{j}\in\mathbb{C}^{n\times1}$
are the transmitted symbol and the signature vector for user $j$
respectively, and $\mathbf{W}\in\mathbb{C}^{n\times1}$ is the additive
white Gaussian noise vector with zero mean and covariance matrix $\sigma^{2}\mathbf{I}$.
The processing gain (length of the signature vector) is $n$, and
$m$ is the number of users. We also assume that the transmitted symbols
$B_{j}$'s are independent and with the same power (variance) $1$. 

We assume that the receiver has perfect knowledge about the signature
vectors $\mathbf{s}_{j}$'s. For a particular user, without loss of
generality, user 1 is assumed, the receiver selects his signature
to avoid the interference from the other users. It feeds the corresponding
index back to user 1 through a finite rate and error-free feedback
link. The rate of the feedback link is assumed to be up to $R_{\mathrm{fb}}$
bits. In order to accomplish this, a signature codebook $\mathcal{B}$
with size $2^{R_{\mathrm{fb}}}$ is declared to both the receiver
and user 1. 

We assume that the signature codebook $\mathcal{B}$ is randomly constructed.
Specifically, $\mathcal{B}=\left\{ \mathbf{v}_{1},\cdots,\mathbf{v}_{2^{R_{\mathrm{fb}}}}\right\} $,
where $\mathbf{v}_{k}=\mathbf{z}_{k}/\left\Vert \mathbf{z}_{k}\right\Vert $,
$\mathbf{z}_{k}=\left[z_{1,k},\cdots,z_{n,k}\right]$ and $z_{i,k}$
are i.i.d. $\mathcal{CN}\left(0,1\right)$ for all $1\leq i\leq n$
and $1\leq k\leq2^{R_{\mathrm{fb}}}$. In this way, it is guaranteed
that $\mathbf{v}_{k}$'s are independent and isotropically distributed
unitary complex vectors. It has been shown that the randomly constructed
codebook is asymptotically optimal \cite{Honig_IT05_CDMA_signature_optimization_finite_feedback,Dai_IT_CDMA_Signature_Optimization}. 

In this paper, we use the average interference as the performance
measure. Let $\mathbf{S}\in\mathbb{C}^{n\times\left(m-1\right)}$
be the interference matrix for user 1, whose columns are the interfering
signatures $\mathbf{s}_{2},\cdots,\mathbf{s}_{m}$. We assume that
$\mathbf{S}$ has i.i.d. complex Gaussian entries with zero mean and
variance $\frac{1}{n}$, same as the assumption in \cite{Honig_IT05_CDMA_signature_optimization_finite_feedback}%
\footnote{It is more natural to assume that the columns in $\mathbf{S}$ are
independent and isotropically distributed unitary complex vectors.
However, the asymptotic statistics of $\mathbf{S}$ are the same for
both assumptions. We adopt the assumption in \cite{Honig_IT05_CDMA_signature_optimization_finite_feedback}
for fair comparison. In Section \ref{sec:Simulations-and-Discussion},
we shall show that the difference between these two assumptions is
indistinguishable for relatively large systems.%
}. For a given interference matrix $\mathbf{S}$, the interference
to user 1 is defined by \[
I_{\mathbf{S}}\triangleq\sum_{j=2}^{m}\left|\left\langle \mathbf{s}_{1},\mathbf{s}_{j}\right\rangle \right|^{2}=\mathbf{s}_{1}^{\dagger}\mathbf{S}\mathbf{S}^{\dagger}\mathbf{s}_{1}.\]
The average interference is defined by $I\triangleq\mathrm{E}_{\mathbf{S}}\left[\mathrm{E}_{\mathcal{B}}\left[I_{\mathbf{S}}\right]\right]$.

In this paper, we consider two types of signature selection criteria.
The first one is to minimize the interference from other users, i.e.,
\begin{equation}
\mathbf{s}_{1}=\underset{\mathbf{v}_{k}\in\mathcal{B}}{\arg\;\min}\;\mathbf{v}_{k}^{\dagger}\mathbf{S}\mathbf{S}^{\dagger}\mathbf{v}_{k}.\label{eq:criterion1}\end{equation}
 The second one is to select the signature vector to match the weakest
interference directions (or equivalently, to be as orthogonal as possible
to the strong interference directions). Let $d$ be the multiplicity
of the smallest singular value of $\mathbf{S}$. Let $\mathbf{u}_{n-d+1},\mathbf{u}_{n-d+2},\cdots,\mathbf{u}_{n}$
be the $d$ left singular vectors of $\mathbf{S}$ corresponding to
the smallest singular value and $\mathbf{U}_{d}=\left[\mathbf{u}_{n-d+1}\cdots\mathbf{u}_{n}\right]$.
The direction matching criterion is \begin{equation}
\mathbf{s}_{1}=\underset{\mathbf{v}_{k}\in\mathcal{B}}{\arg\;\max}\;\mathbf{v}_{k}^{\dagger}\mathbf{U}_{d}\mathbf{U}_{d}^{\dagger}\mathbf{v}_{k}.\label{eq:criterion2}\end{equation}

For both criteria, we shall derive the asymptotic performance limit
in Sections \ref{sec:Interference-Minimization} and \ref{sec:Direction-Match}
respectively. The corresponding analysis can be extended to SINR performance
evaluation for both match filter and linear MMSE receivers \cite{Dai_IT_CDMA_Signature_Optimization}.

\section{\label{sec:Interference-Minimization}Analysis for Interference Minimization}

This section is devoted to calculate the average interference for
the interference minimization criterion in (\ref{eq:criterion1}).
By letting the processing gain, number of users and feedback bits
approach infinity simultaneously with fixed ratios, we derive the
exact performance limit. The result is given in Theorem \ref{thm:Interference-Minimization}. 

\begin{thm}
\label{thm:Interference-Minimization}Define \begin{equation}
d\mu_{\lambda}\triangleq\left\{ \begin{array}{ll}
\frac{\sqrt{\left(\lambda^{+}-\lambda\right)\left(\lambda-\lambda^{-}\right)}}{2\pi\lambda}\mathbbm{1}_{\left[\lambda^{-},\lambda^{+}\right]}d\lambda & \mathrm{if}\; n\leq m\\
\left[\frac{1}{\tau}\frac{\sqrt{\left(\lambda^{+}-\lambda\right)\left(\lambda-\lambda^{-}\right)}}{2\pi\lambda}\mathbbm{1}_{\left[\lambda^{-},\lambda^{+}\right]}\right.\\
\quad\quad\quad\quad\left.+\frac{\tau-1}{\tau}\delta\left(\lambda\right)\right]d\lambda & \mathrm{if}\; n>m\end{array}\right.\label{eq:spectrum-pdf}\end{equation}
 for a $\tau\geq1$, where $\lambda^{\pm}=\left(1\pm\sqrt{\tau}\right)^{2}$.
For convenience, define $\lambda_{t}^{-}\triangleq\lambda^{-}$ if
$n\leq m$ and $\lambda_{t}^{-}\triangleq0$ if $n>m$. For any $x\in\left(\lambda_{t}^{-},\lambda^{+}\right)$
and $\alpha\in\left[0,\frac{1}{x-\lambda_{t}^{-}}\right]$, define\[
\psi\left(x,\alpha\right)\triangleq\int_{\lambda_{t}^{-}}^{\lambda^{+}}\log\left(1+\alpha\left(\lambda-x\right)\right)d\mu_{\lambda}\]
and \[
\bar{\psi}\left(x\right)\triangleq\underset{\alpha\in\left[0,\frac{1}{x-\lambda_{t}^{-}}\right]}{\max}\;\psi\left(x,\alpha\right).\]
Let $n$, $m$ and $R_{\mathrm{fb}}$ approach infinity simultaneously
with fixed ratios $\tau=\max\left(n,m\right)/\min\left(n,m\right)$,
$\bar{r}=\min\left(n,m\right)/n$ and $c=R_{\mathrm{fb}}/n$. For
any $0<c<\infty$, there exists an $x_{c}\in\left(\lambda_{t}^{-},\lambda^{+}\right)$
such that $c\log2=\bar{\psi}\left(x_{c}\right)$ and \begin{equation}
\underset{\left(n,m,R_{\mathrm{fb}}\right)\rightarrow\infty}{\lim}\; I^{\left(n\right)}=\bar{r}x_{c}.\label{eq:Large-Deviation-asymp-result}\end{equation}

\end{thm}

\begin{remrk}
Theorem \ref{thm:Interference-Minimization} is only valid when $0<c=R_{\mathrm{fb}}/n<\infty$.
However, it also provides the exact performance limit when $c\rightarrow0^{+}$
or $c\rightarrow+\infty$. Elementary computations show that as $c\rightarrow0^{+}$,
$\bar{r}x_{c}\rightarrow\bar{r}\bar{\lambda}\triangleq\bar{r}\int_{\lambda_{t}^{-}}^{\lambda^{+}}\lambda d\mu_{\lambda}$
the average eigenvalue, and as $c\rightarrow+\infty$, $\bar{r}x_{c}\rightarrow\bar{r}\lambda_{t}^{-}$
the minimum eigenvalue, which are consistent with intuition.
\end{remrk}

The essential idea behind Theorem \ref{thm:Interference-Minimization}
is the same as that behind the standard large deviation technique.
The lower bound is derived by Chebyshev's inequality and the upper
bound is derived by the twisted distribution. Similar to the result
in large deviation technique, the asymptotic lower and upper bounds
are identical. We shall outline the proofs for the lower and upper
bounds in Section \ref{sub:Proof-LB} and \ref{sub:Proof-UB} respectively.

As a beginning, we express the average interference in a convenient
form. For a CDMA system with finite $n$ and $m$, the average interference
is given by\[
I^{\left(n\right)}=\mathrm{E}_{\mathbf{S}}\left[\mathrm{E}_{\mathcal{B}}\left[\min\;\mathbf{v}_{k}^{\dagger}\mathbf{S}\mathbf{S}^{\dagger}\mathbf{v}_{k}\left|\mathbf{S}\right.\right]\right],\]
where $I^{\left(n\right)}$ is used to emphasize that they are for
finite $n$ and $m$. Let $\mathbf{H}_{n}$ be an $n\times m$ matrix
whose entries are i.i.d. complex Gaussian $\mathcal{CN}\left(0,1\right)$.
Obviously, the statistics of $\mathbf{S}$ is the same as $\frac{1}{\sqrt{n}}\mathbf{H}_{n}$.
Therefore, $\mathbf{S}\mathbf{S}^{\dagger}=\frac{r}{n}\frac{1}{r}\mathbf{H}_{n}\mathbf{H}_{n}^{\dagger}$
where $r\triangleq\min\left(n,m\right)$. Let $\lambda_{i}$ be the
$i^{\mathrm{th}}$ eigenvalue of the matrix $\frac{1}{r}\mathbf{H}_{n}\mathbf{H}_{n}^{\dagger}$.
We have \begin{eqnarray*}
 & I^{\left(n\right)} & \overset{\left(a\right)}{=}\mathrm{E}_{\mathbf{S}}\left[\mathrm{E}_{\mathcal{B}}\left[\underset{k}{\min}\;\frac{\mathbf{z}_{k}^{\dagger}\mathbf{S}\mathbf{S}^{\dagger}\mathbf{z}_{k}}{\left\Vert \mathbf{z}_{k}\right\Vert ^{2}}\left|\mathbf{S}\right.\right]\right]\\
 &  & =\frac{r}{n}\mathrm{E}_{\mathbf{H}_{n}}\left[\mathrm{E}_{\mathcal{B}}\left[\underset{k}{\min}\;\frac{\mathbf{z}_{k}^{\dagger}\frac{1}{r}\mathbf{H}_{n}\mathbf{H}_{n}^{\dagger}\mathbf{z}_{k}}{\left\Vert \mathbf{z}_{k}\right\Vert ^{2}}\left|\mathbf{H}_{n}\right.\right]\right]\\
 &  & \overset{\left(b\right)}{=}\frac{r}{n}\mathrm{E}_{\mathbf{H}_{n}}\left[\mathrm{E}_{\mathcal{B}}\left[\underset{k}{\min}\;\frac{\mathbf{z}_{k}^{\dagger}\mathbf{U\Lambda}\mathbf{U}^{\dagger}\mathbf{z}_{k}}{\left\Vert \mathbf{z}_{k}\right\Vert ^{2}}\left|\mathbf{H}_{n}\right.\right]\right]\\
 &  & \overset{\left(c\right)}{=}\frac{r}{n}\mathrm{E}_{\bm{\lambda}}\left[\mathrm{E}_{\mathcal{B}}\left[\underset{k}{\min}\;\frac{\sum_{i=1}^{n}\lambda_{i}\left|z_{i,k}\right|^{2}}{\sum_{i=1}^{n}\left|z_{i,k}\right|^{2}}\left|\bm{\lambda}\right.\right]\right],\end{eqnarray*}
where 

\begin{description}
\item [(a)]follows from the random construction of the signature codebook
$\mathcal{B}$,
\item [(b)]follows from the singular value decomposition of $\frac{1}{r}\mathbf{H}_{n}\mathbf{H}_{n}^{\dagger}$,
and
\item [(c)]follows from the fact that $\mathbf{z}_{k}$ and $\mathbf{U}\mathbf{z}_{k}$
are statistically equal for any $n\times n$ unitary matrix $\mathbf{U}$
\cite{Muirhead_book82_multivariate_statistics}.
\end{description}
Note that given $\bm{\lambda}$, the random variables $\sum_{i=1}^{n}\lambda_{i}\left|z_{i,k}\right|^{2}/\sum_{i=1}^{n}\left|z_{i,k}\right|^{2}$,
$1\leq k\leq2^{R_{\mathrm{fb}}}$, are i.i.d.. Denote the corresponding
conditional distribution function by $F_{n}\left(x\left|\bm{\lambda}\right.\right)$,
then \begin{eqnarray*}
F_{n}\left(x\left|\bm{\lambda}\right.\right) & = & \Pr\left(\frac{\sum_{i=1}^{n}\lambda_{i}\left|z_{i}\right|^{2}}{\sum_{i=1}^{n}\left|z_{i}\right|^{2}}\leq x\left|\bm{\lambda}\right.\right)\\
 & = & \Pr\left(\sum_{i=1}^{n}\left(\lambda_{i}-x\right)\left|z_{i}\right|^{2}\leq0\left|\bm{\lambda}\right.\right).\end{eqnarray*}
It is worthy to keep in mind that this distribution function is function
of $\bm{\lambda}$. Due to the independence of $\sum_{i=1}^{n}\lambda_{i}\left|z_{i,k}\right|^{2}/\sum_{i=1}^{n}\left|z_{i,k}\right|^{2}$,
$1\leq k\leq2^{R_{\mathrm{fb}}}$, for a given $\bm{\lambda}$, we
have\[
\Pr\left(\underset{k}{\min}\frac{\sum_{i=1}^{n}\lambda_{i}\left|z_{i,k}\right|^{2}}{\sum_{i=1}^{n}\left|z_{i,k}\right|^{2}}\leq x\left|\bm{\lambda}\right.\right)=1-\left(1-F_{n}\left(x\left|\bm{\lambda}\right.\right)\right)^{2^{R_{\mathrm{fb}}}}.\]
Therefore,\begin{eqnarray*}
 &  & \mathrm{E}_{\mathcal{B}}\left[\min\;\frac{\sum_{i=1}^{n}\lambda_{i}\left|z_{j}\right|^{2}}{\sum_{i=1}^{n}\left|z_{j}\right|^{2}}\left|\bm{\lambda}\right.\right]\\
 &  & =\int x\cdot d\left[1-\left(1-F_{n}\left(x\left|\bm{\lambda}\right.\right)\right)^{2^{R_{\mathrm{fb}}}}\right]\\
 &  & =\lambda_{\min}+\int_{\lambda_{\min}}^{\lambda_{\max}}\left(1-F_{n}\left(x\left|\bm{\lambda}\right.\right)\right)^{2^{R_{\mathrm{fb}}}}dx\end{eqnarray*}
and \begin{equation}
I^{\left(n\right)}=\frac{r}{n}\mathrm{E}_{\bm{\lambda}}\left[\lambda_{\min}+\int_{\lambda_{\min}}^{\lambda_{\max}}\left(1-F_{n}\left(x|\bm{\lambda}\right)\right)^{2^{R_{\mathrm{fb}}}}dx\right],\label{eq:I-main-formula}\end{equation}
where $\lambda_{\min}$ and $\lambda_{\max}$ are the minimum and
maximum eigenvalues of $\frac{1}{r}\mathbf{H}_{n}\mathbf{H}_{n}^{\dagger}$
respectively.

It is difficult to calculate (\ref{eq:I-main-formula}) for finite
$n$ and $m$. We let $n$, $m$ and $R_{\mathrm{fb}}$ approach infinity
with fixed ratios and derive lower and upper bounds on $I$ for large
systems.

\subsection{\label{sub:Proof-LB}The Asymptotic Lower Bound}

The following lemma provides an asymptotic lower bound on the average
interference.

\begin{lemma}
\label{lem:Lower-Bound}Following the definitions in Theorem \ref{thm:Interference-Minimization},
let $n$, $m$ and $R_{\mathrm{fb}}$ approach infinity simultaneously
with fixed ratios $\tau$, $\bar{r}$ and $c$. For any $0<c<\infty$,
\[
\underset{\left(n,m,R_{\mathrm{fb}}\right)\rightarrow\infty}{\lim}\; I^{\left(n\right)}\geq\bar{r}x_{c}.\]

\end{lemma}

Due to the length limit, we only sketch the proof. It is based on
Chebyshev's inequality and the asymptotic behavior of the spectrum
of a Wishart matrix. Recall that we are dealing with a distribution
function conditioned on the random vector $\bm{\lambda}$, we need
to define some {}``good'' set of $\bm{\lambda}$, say $A_{\bm{\lambda}}^{n}$,
which will appear soon. By Chebyshev's inequality, it can be proved
that \begin{eqnarray*}
F_{n}\left(x|\bm{\lambda}\right) & = & \Pr\left(\sum_{i=1}^{n}\left(\lambda_{i}-x\right)\left|z_{i}\right|^{2}\leq0\left|\bm{\lambda}\right.\right)\\
 & \leq & \frac{1}{e^{-\alpha\cdot0}}\int e^{-\alpha\sum\left(\lambda_{i}-x\right)\left|z_{i}\right|^{2}}d\mu_{\mathbf{z}}\\
 & = & \exp\left(-\sum_{i=1}^{n}\log\left(1+\alpha\left(\lambda_{i}-x\right)\right)\right)\end{eqnarray*}
 for $\forall\alpha\in\left(0,\frac{1}{x-\lambda_{t}^{-}}\right)$
and $\bm{\lambda}\in A_{\bm{\lambda}}^{n}$, where the set $A_{\bm{\lambda}}^{n}$
is defined by\begin{eqnarray}
A_{\bm{\lambda}}^{n} & \triangleq & \left\{ \bm{\lambda}:\;\left|\psi_{n}\left(\bm{\lambda},x,\alpha\right)-\psi\left(x,\alpha\right)\right|\leq\epsilon_{1}\right\} \nonumber \\
 &  & \cap\left\{ \bm{\lambda}:\;\left|\lambda_{\min}-\lambda_{t}^{-}\right|\leq\epsilon_{2}\right\} ,\label{eq:def-good-set-A}\end{eqnarray}
\[
\psi_{n}\left(\bm{\lambda},x,\alpha\right)\triangleq\frac{1}{n}\sum_{i=1}^{n}\log\left(1+\alpha\left(\lambda_{i}-x\right)\right),\]
and the positive numbers $\epsilon_{1}$ and $\epsilon_{2}$ are small
enough. By the asymptotic behavior of the spectrum of a Wishart matrix,
it can be shown that $\Pr\left(A_{\bm{\lambda}}^{n}\right)\rightarrow1$.

Now take a small $\epsilon>0$ such that $x_{c}-\epsilon>\lambda_{t}^{-}$.
Let $x=x_{c}-\epsilon$. It can be proved that we can always find
an $\alpha\in\left(0,\frac{1}{x-\lambda_{t}^{-}}\right)$ such that
$\psi\left(x,\alpha\right)>c\log2$. Then for $\forall\delta>0$,
\begin{eqnarray}
 &  & \left(1-F_{n}\left(x\left|\bm{\lambda}\right.\right)\right)^{2^{cn}}\nonumber \\
 &  & \geq\left(1-e^{-n\frac{1}{n}\sum\log\left(1+\alpha\left(\lambda_{i}-x\right)\right)}\right)^{2^{cn}}\nonumber \\
 &  & =\exp\left(2^{cn}\log\left(1-e^{-n\frac{1}{n}\sum\log\left(1+\alpha\left(\lambda_{i}-x\right)\right)}\right)\right)\nonumber \\
 &  & \overset{\left(a\right)}{\geq}\exp\left(-e^{-n\left[\psi\left(x,\alpha\right)-\epsilon_{1}-c\log2\right]}\left(1+O\left(1\right)\right)\right)\nonumber \\
 &  & \overset{\left(b\right)}{\geq}1-\delta,\label{eq:lower-bd-extreme-pdf}\end{eqnarray}
on $A_{\bm{\lambda}}^{n}$ for $n$ large enough, where 

\begin{description}
\item [(a)]follows by Taylor series expansion, and
\item [(b)]follows from the fact that we are able to choose $\epsilon_{1}>0$
small enough such that $\psi\left(x,\alpha\right)-\epsilon_{1}-c\log2>0$. 
\end{description}
Then\begin{eqnarray*}
 &  & \mathrm{E}_{\bm{\lambda}}\left[\int_{\lambda_{\min}}^{\lambda_{\max}}\left(1-F_{n}\left(x|\bm{\lambda}\right)\right)^{2^{cn}}dx\right]\\
 &  & \overset{\left(c\right)}{\geq}\mathrm{E}_{\bm{\lambda}}\left[\int_{\lambda_{t}^{-}+\epsilon_{2}}^{x_{c}-\epsilon}\left(1-F_{n}\left(x|\bm{\lambda}\right)\right)^{2^{cn}}dx,\; A_{\bm{\lambda}}^{n}\right]\\
 &  & \overset{\left(d\right)}{\geq}\mathrm{E}_{\bm{\lambda}}\left[\int_{\lambda_{t}^{-}+\epsilon_{2}}^{x_{c}-\epsilon}\left(1-\delta\right)dx,\; A_{\bm{\lambda}}^{n}\right]\\
 &  & =\left(1-\delta\right)\left(x_{c}-\lambda_{t}^{-}-\epsilon-\epsilon_{2}\right)\cdot\Pr\left(A_{\bm{\lambda}}^{n}\right)\\
 &  & \rightarrow\left(1-\delta\right)\left(x_{c}-\lambda_{t}^{-}-\epsilon-\epsilon_{2}\right),\end{eqnarray*}
where 

\begin{description}
\item [(c)]follows by reducing the integration domain of a non-negative
function, and
\item [(d)]follows from (\ref{eq:lower-bd-extreme-pdf}) and the fact that
$F_{n}\left(x|\bm{\lambda}\right)$ is a non-decreasing function in
$x$.
\end{description}
Therefore, \begin{eqnarray*}
\lim I^{\left(n\right)} & = & \bar{r}\lim\mathrm{E}_{\bm{\lambda}}\left[\lambda_{\min}+\int_{\lambda_{\min}}^{\lambda_{\max}}\left(1-F_{n}\left(x|\bm{\lambda}\right)\right)^{2^{cn}}dx\right]\\
 & \geq & \bar{r}\left[\lambda_{t}^{-}+\left(1-\delta\right)\left(x_{c}-\lambda_{t}^{-}-\epsilon-\epsilon_{2}\right)\right].\end{eqnarray*}
By taking $\delta$, $\epsilon$ and $\epsilon_{2}$ arbitrarily small,
we have $\lim I^{\left(n\right)}\geq\bar{r}x_{c}$.

\subsection{\label{sub:Proof-UB}The Asymptotic Upper Bound}

For the interference minimization criterion in (\ref{eq:criterion1}),
the asymptotic upper bound on the average interference is given in
Lemma \ref{lem:Upper-Bound}.

\begin{lemma}
\label{lem:Upper-Bound}Following the definitions in Theorem \ref{thm:Interference-Minimization},
let $n$, $m$ and $R_{\mathrm{fb}}$ approach infinity simultaneously
with fixed ratios $\tau$, $\bar{r}$ and $c$. For any $0<c<\infty$,
\[
\underset{\left(n,m,R_{\mathrm{fb}}\right)\rightarrow\infty}{\lim}\; I^{\left(n\right)}\leq\bar{r}x_{c}.\]

\end{lemma}

Due to the length limit, we omit the detailed proof. A sketch of the
proof is given in the below.

To prove the upper bound, roughly speaking, it is sufficient to show
that for $\forall\epsilon>0$ and $\forall\delta>0$, if $n$ is large
enough, we can upper bound $\left(1-F_{n}\left(x|\bm{\lambda}\right)\right)^{2^{cn}}$
uniformly by $\delta$, i.e.,\begin{equation}
\left(1-F_{n}\left(x|\bm{\lambda}\right)\right)^{2^{cn}}<\delta\;\mathrm{for\; all}\; x>x_{c}+\epsilon,\label{eq:upper-bd-extreme-pdf}\end{equation}
on the {}``good'' set $A_{\bm{\lambda}}^{n}$ (\ref{eq:def-good-set-A}).
Since $\Pr\left(A_{\bm{\lambda}}^{n}\right)\rightarrow1$, \begin{eqnarray*}
I^{\left(n\right)} & = & \mathrm{E}_{\bm{\lambda}}\left[\lambda_{\min}+\int_{\lambda_{\min}}^{\lambda_{\max}}\left(1-F_{n}\left(x|\bm{\lambda}\right)\right)^{2^{R_{\mathrm{fb}}}}dx\right]\\
 & \leq & \bar{r}\left\{ \mathrm{E}_{\bm{\lambda}}\left[\lambda_{\min}\right]+\mathrm{E}_{\bm{\lambda}}\left[\int_{\lambda_{\min}}^{x_{c}+\epsilon}1dx\right]\right.\\
 &  & \quad\quad+\mathrm{E}_{\bm{\lambda}}\left[\int_{x_{c}+\epsilon}^{\lambda_{\max}}1dx,\Omega_{\bm{\lambda}}-A_{\bm{\lambda}}^{n}\right]\\
 &  & \quad\quad\left.+\mathrm{E}_{\bm{\lambda}}\left[\int_{x_{c}+\epsilon}^{\lambda_{\max}}\delta dx,A_{\bm{\lambda}}^{n}\right]\right\} \\
 & = & \bar{r}\left\{ \mathrm{E}_{\bm{\lambda}}\left[x_{c}+\epsilon\right]+\left(\lambda_{\max}-x_{c}-\epsilon\right)\left(1-\Pr\left(A_{\bm{\lambda}}^{n}\right)\right)\right.\\
 &  & \quad\quad\left.+\delta\left(\lambda_{\max}-x_{c}-\epsilon\right)\Pr\left(A_{\bm{\lambda}}^{n}\right)\right\} \\
 & \rightarrow & \bar{r}\left\{ x_{c}+\epsilon+\delta\left(\lambda_{\max}-x_{c}-\epsilon\right)\right\} .\end{eqnarray*}
By taking $\epsilon$ and $\delta$ arbitrarily small, we have \[
\lim I^{\left(n\right)}\leq\bar{r}x_{c}.\]

The essential tool used to prove (\ref{eq:upper-bd-extreme-pdf})
is the \textbf{twisted distribution} \cite{Dembo_book92_Large_Deviation}.
This tool is the main tool in proving the lower bound of Cramer's
theorem \cite{Dembo_book92_Large_Deviation}, a basic result of large
deviations. However, there is a fundamental difference between the
standard large deviation technique and our approach. While in Cramer's
theorem one considers the sums of i.i.d. random variables, here we
consider the sum of $\left(\lambda_{i}-x\right)\left|z_{i}\right|^{2}$,
where the random variables are conditional independent but not identically
distributed and the condition itself is a random vector. While the
conditional independence requires us to discuss the statistics of
$\sum\left(\lambda_{i}-x\right)\left|z_{i}\right|^{2}$ on the set
$A_{\bm{\lambda}}^{n}$, the non-identical distribution brings the
major difficulty. That is, the twisted distribution may or may not
be well-defined. To overcome this difficulty, we have to discuss two
types of $x$ and define two types of twisted distributions respectively.

We define two types of $x$ and two types of twisted distributions
as follows. Let $\alpha_{x}$ be the $\alpha$ such that $\bar{\psi}\left(x\right)=\psi\left(x,\alpha_{x}\right)$.
The set of $x$ of the first type is defined by \[
\mathcal{X}_{1}\triangleq\left\{ x\in\left(\lambda_{t}^{-},\lambda^{+}\right):\;\alpha_{x}\in\left(0,\frac{1}{x-\lambda_{t}^{-}}\right)\right\} .\]
The set of $x$ of the second type is defined by \[
\mathcal{X}_{2}\triangleq\left\{ x\in\left(\lambda_{t}^{-},\lambda^{+}\right):\;\alpha_{x}=\frac{1}{x-\lambda_{t}^{-}}\right\} .\]
It can be proved that the $x_{c}$ in Theorem \ref{thm:Interference-Minimization}
is either in $\mathcal{X}_{1}$ or in $\mathcal{X}_{2}$. If $x\in\mathcal{X}_{1}$,
then $\alpha_{x}<\frac{1}{x-\lambda_{t}^{-}}$ and $\mathrm{E}_{\mathbf{z}}\left[e^{-\alpha\sum\left(\lambda_{i}-x\right)\left|z_{i}\right|^{2}}\right]$
is well defined on the set $A_{\bm{\lambda}}^{n}$ with $\epsilon_{2}$
small enough. Then we are able to define a twisted distribution measure\[
d\tilde{\mu}_{\mathbf{z}}\triangleq\frac{e^{-\alpha\sum\left(\lambda_{i}-x\right)\left|z_{i}\right|^{2}}}{\mathrm{E}_{\mathbf{z}}\left[e^{-\alpha\sum\left(\lambda_{i}-x\right)\left|z_{i}\right|^{2}}\right]}d\mu_{\mathbf{z}},\]
where $d\mu_{\mathbf{z}}$ is the probability measure for the random
vector $\mathbf{z}$. However, if $x\in\mathcal{X}_{2}$, $\mathrm{E}_{\mathbf{z}}\left[e^{-\alpha\sum\left(\lambda_{i}-x\right)\left|z_{i}\right|^{2}}\right]$
is not well defined on the set $A_{\bm{\lambda}}^{n}$ no matter how
small $\epsilon_{2}>0$ we choose. For this case, we have to define
the twisted distribution in a {}``truncated'' way. Define the $M$-truncated
measure for $\mathbf{z}$ as \[
d\mu_{\mathbf{z}}^{M}=\prod_{i=1}^{n}\mathbbm{1}_{z_{i}\in\left[0,M\right]}d\mu_{\mathbf{z}}.\]
Then the $M$-truncated twisted distribution measure is defined by\[
d\tilde{\mu}_{\mathbf{z}}^{M}\triangleq\frac{e^{-\alpha\sum\left(\lambda_{i}-x\right)\left|z_{i}\right|^{2}}}{\mathrm{E}_{\mu_{\mathbf{z}}^{M}}\left[e^{-\alpha\sum\left(\lambda_{i}-x\right)\left|z_{i}\right|^{2}}\right]}d\mu_{\mathbf{z}}^{M}.\]
It can be verified that $d\tilde{\mu}_{\mathbf{z}}^{M}$ is always
well defined on the set $A_{\bm{\lambda}}^{n}$ for a finite $M>0$.
In the proof of (\ref{eq:upper-bd-extreme-pdf}), we need to choose
an $M$ sufficiently large.

With the twisted distributions, (\ref{eq:upper-bd-extreme-pdf}) can
be proved. Here, we only outline the proof for $x_{c}\in\mathcal{X}_{1}$,
the simpler case. Assume that $x_{c}\in\mathcal{X}_{1}$. For an $\epsilon>0$
small enough, let $x=x_{c}+\frac{\epsilon}{2}$. It can be proved
that $x\in\mathcal{X}_{1}$ and $\bar{\psi}\left(x\right)=\psi\left(x,\alpha_{x}\right)<c\log2$.
For $\forall\delta_{1}>0$ and a $y>x=x_{c}+\frac{\epsilon}{2}$,
it can be proved that \[
F_{n}\left(y|\bm{\lambda}\right)\geq\Pr\left(\sum\left(\lambda_{i}-x\right)\left|z_{i}\right|^{2}\leq n\epsilon_{3}\right)\left(1-\delta_{1}\right)\]
 on the set $A_{\bm{\lambda}}^{n}$ with small enough $\epsilon_{3}$
and large enough $n$. But for any given $\epsilon_{3}>0$, a further
lower bound can be derived as follows. \begin{eqnarray*}
 &  & \Pr\left(\sum\left(\lambda_{i}-x\right)\left|z_{i}\right|^{2}\leq n\epsilon_{3}\right)\\
 & \overset{\left(a\right)}{\geq} & \Pr\left(-\alpha_{x}\sum\left(\lambda_{i}-x\right)\left|z_{i}\right|^{2}\in n\left(-\delta_{2},\delta_{2}\right)\right)\\
 & \overset{\left(b\right)}{\geq} & e^{-n\delta_{2}}\int_{B_{\mathbf{z}}}e^{-\alpha_{x}\sum\left(\lambda_{i}-x\right)\left|z_{i}\right|^{2}}d\mu_{\mathbf{z}}\\
 & \overset{\left(c\right)}{=} & e^{-n\delta_{2}}\mathrm{E}_{\mathbf{z}}\left[e^{-\alpha_{x}\sum\left(\lambda_{i}-x\right)\left|z_{i}\right|^{2}}\right]\int_{B_{\mathbf{z}}}d\tilde{\mu}_{\mathbf{z}}\\
 & \overset{\left(d\right)}{\geq} & \exp\left\{ -n\left[\delta'+\psi\left(x,\alpha_{x}\right)\right]\right\} P_{\tilde{\mu}_{\mathbf{z}}}\left(B_{\mathbf{z}}\right),\end{eqnarray*}
on the set $A_{\bm{\lambda}}^{n}$ with large enough $n$, where 

\begin{description}
\item [(a)]holds by choosing $\delta_{2}<\frac{\epsilon}{\alpha_{x}}$,
\item [(b)]holds by defining \[
B_{\mathbf{z}}\triangleq\left\{ \mathbf{z}:\;-\alpha_{x}\sum\left(\lambda_{i}-x\right)\left|z_{i}\right|^{2}\in n\left(-\delta_{2},\delta_{2}\right)\right\} ,\]

\item [(c)]follows from the definition of the twisted distribution for
$x\in\mathcal{X}_{1}$, and
\item [(d)]follows by letting $\delta'=\epsilon_{1}+\delta_{2}$, where
$P_{\tilde{\mu}_{\mathbf{z}}}\left(B_{\mathbf{z}}\right)$ is the
probability of $B_{\mathbf{z}}$ under the twisted distribution. 
\end{description}
We want to calculate $P_{\tilde{\mu}_{\mathbf{z}}}\left(B_{\mathbf{z}}\right)$.
By studying the asymptotic behavior of $\frac{d}{d\alpha}\log\;\mathrm{E}_{\mathbf{z}}\left[e^{-\alpha_{x}\sum\left(\lambda_{i}-x\right)\left|z_{i}\right|^{2}}\right]$,
it can be proved that for $\forall\delta''>0$, $P_{\tilde{\mu}_{\mathbf{z}}}\left(B_{\mathbf{z}}\right)\geq1-\delta''$
on the set $A_{\bm{\lambda}}^{n}$ with large enough $n$. Therefore,
for $\forall\delta'''>0$, \[
F_{n}\left(y|\bm{\lambda}\right)\geq e^{-n\left[\bar{\psi}\left(x\right)+\delta'\right]}\left(1-\delta'''\right)\]
 on the set $A_{\bm{\lambda}}^{n}$ with large enough $n$. Now we
choose $\delta'$ small enough such that $\bar{\psi}\left(x\right)+2\delta'<c\log2$.
Then it can be proved that, for $\forall\delta>0$, \begin{eqnarray*}
 &  & \left[1-F_{n}\left(y|\bm{\lambda}\right)\right]^{2^{cn}}\\
 & \leq & \exp\left[-\left(1-\delta'''\right)e^{-n\left[\bar{\psi}\left(x\right)+\delta'-c\log2\right]}\right]\\
 & < & \delta\end{eqnarray*}
on the set $A_{\bm{\lambda}}^{n}$ with large enough $n$. Without
loss of generality, we take $y=x_{c}+\epsilon>x=x_{c}+\frac{\epsilon}{2}$.
Since $F_{n}\left(y|\bm{\lambda}\right)$ is a non-decreasing function,
we have uniform boundedness, \[
\left[1-F_{n}\left(y|\bm{\lambda}\right)\right]^{2^{cn}}<\delta\;\mathrm{for\; all}\; y>x_{c}+\epsilon\]
on the set $A_{\bm{\lambda}}^{n}$ with large enough $n$. This is
(\ref{eq:upper-bd-extreme-pdf}), what we want.

\section{\label{sec:Direction-Match} Direction Matching Criterion}

In this section, we shall analyze the performance corresponding to
the direction matching criterion in (\ref{eq:criterion2}). Again,
by letting $n$, $m$ and $R_{\mathrm{fb}}$ approach infinity simultaneously
with fixed ratios, we derive the exact performance limit. The result
is given in Theorem \ref{thm:Direction-Matching}.

\begin{thm}
\label{thm:Direction-Matching}Following the definitions in Theorem
\ref{thm:Interference-Minimization}, let $n$, $m$ and $R_{\mathrm{fb}}$
approach infinity simultaneously with fixed ratios $\tau$, $\bar{r}$
and $c$. For any $0<c<\infty$, \begin{equation}
\underset{\left(n,m,R_{\mathrm{fb}}\right)\rightarrow\infty}{\lim}I^{\left(n\right)}=\left\{ \begin{array}{ll}
\lambda_{t}^{-}\left(1-2^{-c}\right)+\bar{\lambda}2^{-c} & \mathrm{if}\; n\leq m\\
x_{c} & \mathrm{if}\; n>m\end{array}\right.,\label{eq:Grassmann-asymp-result}\end{equation}
where $\bar{\lambda}=\frac{m}{n}$, $x_{c}<\frac{m}{n}$ satisfies
$D\left(\mu_{\bar{r}}\parallel\mu_{x_{c}}\right)=c\log2$ and $D\left(\mu_{\bar{r}}\parallel\mu_{x_{c}}\right)\triangleq\bar{r}\log\frac{\bar{r}}{x_{c}}+\left(1-\bar{r}\right)\log\frac{1-\bar{r}}{1-x_{c}}$
is known as the relative entropy.
\end{thm}

\begin{remrk}
Elementary calculations show that the asymptotic average interference
$\lim\; I$, as a function of $c$, converges to the average eigenvalue
and the minimum eigenvalue as $c\rightarrow0^{+}$ and $c\rightarrow\infty$
respectively. These results are consistent with intuitions.
\end{remrk}

The proof of Theorem \ref{thm:Direction-Matching} is based on the
observation that \begin{eqnarray*}
I^{\left(n\right)} & = & \bar{r}\sum_{i=1}^{n-d}\mathrm{E}_{\mathbf{U}}\left[\mathrm{E}_{\mathcal{B}}\left[\mathbf{s}_{1}^{\dagger}\mathbf{u}_{i}\mathbf{u}_{i}^{\dagger}\mathbf{s}_{1}\right]\right]\mathrm{E}_{\mathbf{\Lambda}}\left[\lambda_{i}\right]\\
 &  & \quad\;+\bar{r}\sum_{i=n-d+1}^{n}\mathrm{E}_{\mathbf{U}}\left[\mathrm{E}_{\mathcal{B}}\left[\mathbf{s}_{1}^{\dagger}\mathbf{u}_{i}\mathbf{u}_{i}^{\dagger}\mathbf{s}_{1}\right]\right]\mathrm{E}_{\mathbf{\Lambda}}\left[\lambda_{n}\right].\end{eqnarray*}
where $\lambda_{1}\geq\cdots\geq\lambda_{n}$ are the singular values
of $\frac{1}{r}\mathbf{H}_{n}\mathbf{H}_{n}^{\dagger}$ and $\mathbf{u}_{i}$
is the singular vector corresponding $\lambda_{i}$. For $n\leq m$
(full rank) case, $d=1$ with probability 1. We select the signature
$\mathbf{s}_{1}$ to match $\mathbf{u}_{n}$. The corresponding $\mathrm{E}_{\mathbf{U}}\left[\mathrm{E}_{\mathcal{B}}\left[\mathbf{s}_{1}^{\dagger}\mathbf{u}_{n}\mathbf{u}_{n}^{\dagger}\mathbf{s}_{1}\right]\right]$
can be calculated based on our previous results in the Grassmann manifold
\cite{Dai_Globecom05_Quantization_bounds_Grassmann_manifold}. For
$n>m$ (deficient rank) case, $d=n-m$ with probability 1. We need
to choose the signature $\mathbf{s}_{1}$ to match the plane generated
by $\mathbf{U}_{d}=\left[\mathbf{u}_{n-d+1}\cdots\mathbf{u}_{n}\right]$.
By large deviation technique, the corresponding $\mathrm{E}_{\mathbf{U}}\left[\mathrm{E}_{\mathcal{B}}\left[\mathbf{s}_{1}^{\dagger}\mathbf{U}_{d}\mathbf{U}_{d}^{\dagger}\mathbf{s}_{1}\right]\right]$
can be evaluated. The detailed derivation is given in \cite{Dai_IT_CDMA_Signature_Optimization}.

\section{\label{sec:Simulations-and-Discussion}Simulations and Discussion}

\begin{figure}
\subfigure[Full Rank ($n \leq m$) Case ]{\includegraphics[%
  clip,
  scale=0.55]{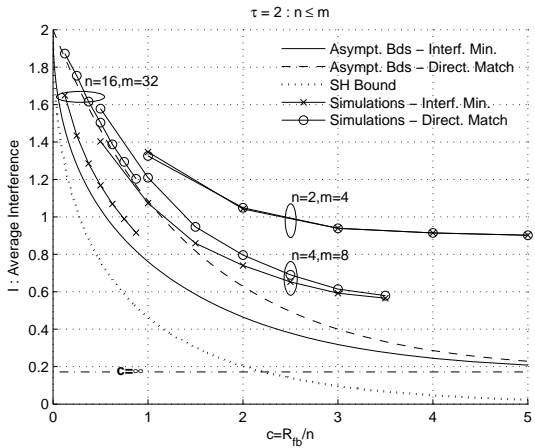}}

\subfigure[Deficient Rank ($n>m$) Case]{\includegraphics[%
  clip,
  scale=0.55]{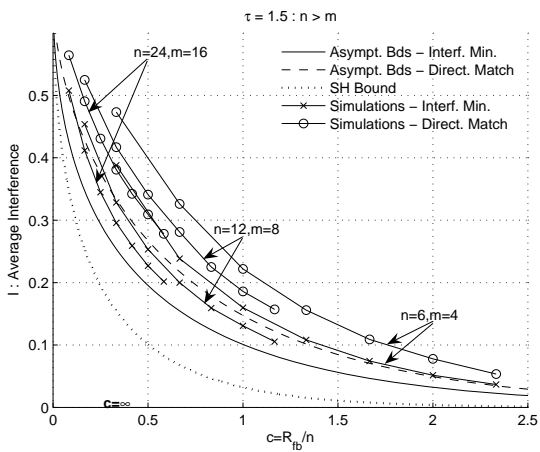}}

\caption{\label{cap:fig-criteria}Simulations for both signature selection
criteria}
\end{figure}
Fig. \ref{cap:fig-criteria} shows the simulation results to demonstrate
the asymptotic performance formulas (\ref{eq:Large-Deviation-asymp-result})
and (\ref{eq:Grassmann-asymp-result}) for both criteria. Fig. \ref{cap:fig-criteria}(a)
and \ref{cap:fig-criteria}(b) are for $n\leq m$ (full rank) and
$n>m$ (deficient rank) cases respectively. From the simulations,
we can observe that the simulated average interference for both criteria
(x markers for interference minimization and circles for direction
matching) converges to the asymptotic results (the solid line for
interference minimization and the dashed line for direction matching)
as $n$, $m$ and $R_{\mathrm{fb}}$ approach infinity with fixed
ratios. Simulations also show that direction matching is a sub-optimal
criterion.

We also compare our formula with the bound in \cite{Honig_IT05_CDMA_signature_optimization_finite_feedback},
denoted as SH bound in Fig \ref{cap:fig-criteria}. In \cite{Honig_IT05_CDMA_signature_optimization_finite_feedback},
an asymptotic lower bound on the average interference is given for
the interference minimization criterion. It is plotted as the dotted
line in Fig. \ref{cap:fig-criteria}. Note that when $\tau=2$ and
$n\leq m$, with infinite feedback rate ($c=\infty$), $I^{\left(n\right)}$
should converge to the minimum eigenvalue $\left(1-\sqrt{2}\right)^{2}\approx0.17$.
The bound in \cite{Honig_IT05_CDMA_signature_optimization_finite_feedback}
is below this value even when $c$ is relatively small ($c\geq2.5$
in Fig \ref{cap:fig-criteria}). Generally speaking, the bound in
\cite{Honig_IT05_CDMA_signature_optimization_finite_feedback} under-estimates
the interference while our asymptotic formula (\ref{eq:Large-Deviation-asymp-result})
gives the exact performance limit.

\begin{figure}
\includegraphics[%
  clip,
  scale=0.55]{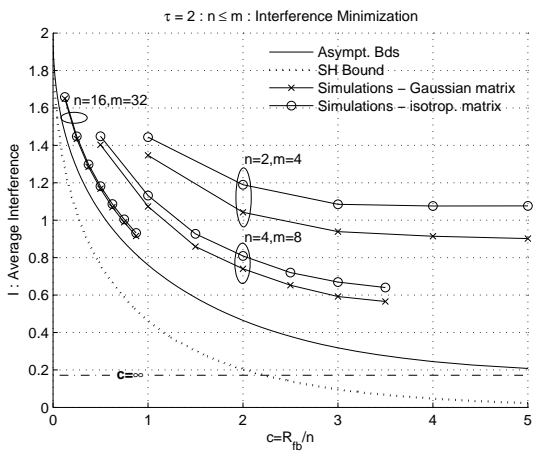}

\caption{\label{cap:fig-random-matrices}Comparison of two types of interference
matrices}
\end{figure}

As mentioned in the system model section, we assume that the interference
matrix $\mathbf{S}$ has i.i.d. complex Gaussian entries with zero
mean and variance $\frac{1}{n}$ for fair comparison, while it is
more natural to assume that $\mathbf{S}$ has independent and isotropically
distributed unitary complex columns. Fig \ref{cap:fig-random-matrices}
gives the difference between these two statistical assumptions, where
the simulations are based on the interference minimization criterion.
For small $n$ and $m$, these two different assumptions give two
different results. However, as $n$ and $m$ increase, for example,
$n=16$ and $m=32$, the difference becomes indistinguishable. Indeed,
the asymptotic statistics of these two types of random matrices are
identical. The results (\ref{eq:Large-Deviation-asymp-result}) and
(\ref{eq:Grassmann-asymp-result}) are the exact asymptotic performance
for both interference statistical assumptions.

\section{\label{sec:Conclusion}Conclusion}

In this paper, we quantify the average interference as a function
of finite feedback rate for CDMA signature optimization problem. Two
signature selection criteria, i.e., interference minimization and
direction matching, are analyzed. By letting the processing gain,
number of users and feedback bits approach infinity with fixed ratios,
we derive the exact asymptotic formulas to calculate the average interference
for both criteria respectively. The asymptotic results are valid for
both the Gaussian interference matrix and the interference matrix
with independent and isotropically distributed columns. Furthermore,
the corresponding analysis can be extended to SINR performance evaluation
for both match filter and linear MMSE receivers.

\bibliographystyle{IEEEtran}
\bibliography{bib/_Liu_Dai,bib/Honig,bib/RandomMatrix}

\begin{thebibliography}{1}
\providecommand{\url}[1]{#1}
\csname url@rmstyle\endcsname
\providecommand{\newblock}{\relax}
\providecommand{\bibinfo}[2]{#2}
\providecommand\BIBentrySTDinterwordspacing{\spaceskip=0pt\relax}
\providecommand\BIBentryALTinterwordstretchfactor{4}
\providecommand\BIBentryALTinterwordspacing{\spaceskip=\fontdimen2\font plus
\BIBentryALTinterwordstretchfactor\fontdimen3\font minus
  \fontdimen4\font\relax}
\providecommand\BIBforeignlanguage[2]{{%
\expandafter\ifx\csname l@#1\endcsname\relax
\typeout{** WARNING: IEEEtran.bst: No hyphenation pattern has been}%
\typeout{** loaded for the language `#1'. Using the pattern for}%
\typeout{** the default language instead.}%
\else
\language=\csname l@#1\endcsname
\fi
#2}}

\bibitem{Honig_IT05_CDMA_signature_optimization_finite_feedback}
W.~Santipach and M.~Honig, ``Signature optimization for {CDMA} with limited
  feedback,'' \emph{Information Theory, IEEE Transactions on}, vol.~51, no.~10,
  pp. 3475--3492, 2005.

\bibitem{Dai_IT_CDMA_Signature_Optimization}
W.~Dai, Y.~Liu, and B.~Rider, ``On the performance of {CDMA} signature
  optimization with finite rate feedback,'' In preparation for journal
  submission.

\bibitem{Muirhead_book82_multivariate_statistics}
R.~J. Muirhead, \emph{Aspects of multivariate statistical theory}.\hskip 1em
  plus 0.5em minus 0.4em\relax New York: John Wiley and Sons, 1982.

\bibitem{Dembo_book92_Large_Deviation}
A.~Dembo and O.~Zeitouni, \emph{Large Deviation Techniques in Decision,
  Simulation, and Estimation}.\hskip 1em plus 0.5em minus 0.4em\relax Jones and
  Bartlett Publishers, 1992.

\bibitem{Dai_Globecom05_Quantization_bounds_Grassmann_manifold}
W.~Dai, Y.~Liu, and B.~Rider, ``Quantization bounds on {G}rassmann manifolds of
  arbitrary dimensions and {MIMO} communications with feedback,'' in \emph{IEEE
  Global Telecommunications Conference (GLOBECOM)}, 2005.

\end{thebibliography}

\end{document}